\documentclass[manuscript]{aastex63}

\submitjournal{ApJ}

\shorttitle{Recurrent galactic cosmic-ray flux modulations}
\shortauthors{Catia Grimani}
\graphicspath{{./}{figures/}}
\begin{document}

\title{Recurrent galactic cosmic-ray flux modulation in L1 and geomagnetic activity during the declining phase of the solar cycle 24}

\correspondingauthor{catia.grimani@uniurb.it}
\email{catia.grimani@uniurb.it}

\author[0000-0002-5467-6386]{Catia Grimani}
\affiliation{DiSPeA, University of Urbino Carlo Bo, Urbino (PU), Italy}
\affiliation{INFN, Florence, Italy}

\author[0000-0002-8611-8610]{Andrea Cesarini}
\affiliation{INFN, Florence, Italy}

\author[0000-0002-2464-1369]{Michele Fabi}
\affiliation{DiSPeA, University of Urbino Carlo Bo, Urbino (PU), Italy}
\affiliation{INFN, Florence, Italy}

\author{Federico Sabbatini}
\affiliation{Alma Mater University, Bologna, Italy}

\author[0000-0002-6710-8142]{Daniele Telloni}
\affiliation{National Institute of Astrophysics, Astrophysical Observatory of Turin, Pino Torinese, Italy}

\author[0000-0003-2429-1626]{Mattia Villani}
\affiliation{DiSPeA, University of Urbino Carlo Bo, Urbino (PU), Italy}
\affiliation{INFN, Florence, Italy}



\begin{abstract}

Galactic cosmic-ray (GCR) flux short-term variations ($<$1 month) in the inner heliosphere are mainly associated with the passage of high-speed solar wind streams (HSS) and interplanetary (IP) counterparts of coronal mass ejections (ICMEs). Data gathered with a particle detector flown on board the ESA LISA Pathfinder (LPF) spacecraft, during the declining part of the solar cycle 24 (February 2016 - July 2017) around the Lagrange point L1, have allowed to study the characteristics of  recurrent cosmic-ray flux modulations above 70 MeV n$^{-1}$. 
%These modulations are observed when the solar wind speed is $>$ 400 km s$^{-1}$ and/or the IP magnetic field intensity $>$ 10 nT. 
It is shown that the amplitude and evolution of individual  modulations depend in a unique way on both IP plasma parameters and particle flux intensity before HSS and ICMEs transit. By comparing the LPF data with those gathered contemporaneously with the magnetic spectrometer experiment AMS-02 on board the International Space Station and with those of Earth polar neutron monitors, the GCR flux modulation was studied at different energies during recurrent short-term variations. It is also aimed  to set the near real-time particle observation requirements to disentangle  the role of long and short-term variations of the GCR flux to evaluate the performance of high-sensitivity instruments in space such as the future interferometers for gravitational wave detection. Finally, the association between recurrent GCR flux variation  observations in L1 and weak to moderate geomagnetic activity in 2016-2017 is discussed.  Short-term recurrent GCR flux variations are  good proxies of recurrent geomagnetic activity when the B$_z$ component of the IP magnetic field is directed northern.

\end{abstract}



\section{Introduction} \label{sec:intro}

The galactic cosmic-ray (GCR) flux at the interstellar medium differs from observations gathered in the inner heliosphere due to a modulation process following the  particle  propagation throught the heliosheath, heliopause and the heliosphere \citep{florinski}. The GCR flux shows a modulation varying with the 11-year solar cycle and the 22-year polarity reversal of the global solar magnetic field (GSMF). This modulation is ascribable to the particle propagation against the outward motion of  the solar wind and embedded magnetic field \citep{balogh}. Cosmic rays diffuse, scatter and  drift through the solar wind and across inhomogeineties of the interplanetary (IP) magnetic field \citep{jokipii}.
 Consequently, the cosmic-ray flux modulation in the inner heliosphere appears  space, time, charge and energy dependent. Particles with energies below tens of MeV are convected outward before reaching the inner heliosphere while any difference between IP and interstellar GCR energy spectra   vanishes above tens of GeV \citep{pogorelov}.  
%Between  70 MeV and tens of GeV particle energy losses depend on the speed of the solar wind, the heliospheric magnetic field configuration, intensity  and turbulence and therefore vary with the solar cycle. 
The  GCR  proton integral flux  above 70 MeV is also observed to decrease by approximately a factor of four with time from solar minimum through  solar maximum at 1 AU \citep{apj1}. Spatial variations studied with the PAMELA and Ulysses experiments, for instance,  revealed that the GCR intensity increases of 2.7\% AU$^{-1}$ with  increasing  radial distance from the Sun and decreases of (0.024$\pm$ 0.005)\% deg$^{-1}$ with increasing heliolatitude 
 \citep{desimone}. In addition to long-term GCR flux variations,  
%modulation  originates in the solar wind  and imbebbed magnetic field, it is observed to  
%decreases with distance from the Sun and the heliolatitude (refs). 
%In this  work  it is focused on cosmic-ray observations gathered  during the descending phase of the solar cycle 24 around the Lagrange point L1.
% variations with time can be disentangled from space variations.
the passage of  high-speed solar wind streams (HSS), IP counterparts of coronal mass ejections (ICMEs) and  heliospheric current sheet crossing (HCSC) were observed to cause short-term modulations  of the overall cosmic-ray flux intensity  varying from a few percent to  tens percent \citep{mccracken,apj1,apj2}. In particular, during the declining phase of the solar cycle,  solar wind streams with speeds  $>>$ 400 km s$^{-1}$, originating from near equatorial coronal holes and equatorward extensions of polar coronal holes, are observed at the ecliptic plane \citep{tsurutani}. HSS generated from coronal holes  lasting more than one solar rotation,  appear to corotate with the Sun. Fast and slow upstream wind interact generating regions of high-magnetic field compressed plasma called corotating interaction regions (CIRs; \citet{richardson1}, \citet{apj1} and references therein).   
%above 70 MeV n$^{-1}$. 
%  the eleven year solar cycle and the twenty-two year polarity reversal of the global solar magnetic field (GSMF).
The total residence time of cosmic rays in the outer and inner heliosphere is strongly energy dependent and result from longer residence time in the heliotail and heliosheath with respect to that in the heliosphere. The residence time in the heliosphere, in particular, ranges between hundreds of days at 100 MeV n$^{-1}$ through tens of days above 1 GeV \citep{pogorelov}.  During this time galactic particles  go through HSS and structures in a disturbed IP medium. 
%In general, when cosmic-ray propagation models in the inner heliosphere  are compared to observations, model parameters are set by considering  the long-term modulation only. 
%of cosmic rays, while unavoidibly observations include the effects of both long and short variations. 
The most precise data on GCR differential fluxes  used to test the particle propagation models in the inner heliosphere  are those gathered with  magnetic spectrometer experiments consisting  of several detectors for redundant particle species selection and energy measurement (see for instance \cite{corti}). 
%Unfortunately,
% these experiments are   characterized by small geometrical factors, therefore,  in order to reduce the statistical uncertainties,  
Energy differential fluxes from a few tens of MeV through hundreds of GeV measured by the PAMELA and AMS-02 experiments are reported in   \citet{martucci,aguilar2018}. Data gathered monthly are affected by both long and short term modulations \citep{pamAPJ,pamAPJL,pamICRC}. 
% and only for a limited amount of time the flux of cosmic rays result only from particle propagation from the interstellar medium to the point of 
%observation in an undisturbed interplenetary medium. 
%The reliability of models for interplanetary propagation of cosmic rays are verified on the basis of space magnetic spectrometer experiments providing precise energy differential flux observations. Unfortunately, however these experiments are characterized by small geometrical factors and, in general, data are integrated over moths period, thus including the effects of both average solar activity and interplanetary propagating structures. 
%This implies that when models of GCR flux propagation are  compared to observations to set montly the values of the solar modulation parameter (ref), it must be taken into account that observations are affected by both long and short-term variations. 
%different periods of the solar cycle are very differently affected by interplanetary structures and that database reporting solar modulation parameters, include in this parameter the undistinguished effects of long and short-term variations.
%In this work an approach applied to non-recurrent Forbush decreases, is proposed to be considered to study the energy dependence of short-term variations
%if the GCR flux depend on the phase of the solar cycle. 
In particular, recurrent  short-term variations appear mainly correlated with   HSS transit when cosmic-ray particles undergo diffusion from enhanced IP magnetic field and convection in the solar wind  \citep{sabbah}. Long and
% and equatorward extensions of polar coronal holes observed mainly during the descending phase of  solar cycles \cite{tsurutani}.
 short-term  GCR flux variations are  observed to be energy dependent with particle spectra being  more depressed at hundreds of MeV with respect to GeV energies. Moreover, short-term modulation of the GCR flux is strongly space dependent  \citep{simoapj}.
%at the transit of IP structures.
%is found in the majority of cases to be at the origin of GCR flux recurrent variations (refs) coronal holes.
%The passage  of HSS is mainly observed during the descending phase of  the solar cycle due to equatorial and equatorward extensions of polar coronal holes.
%, while interplanetary counterparts of coronal mass ejections are mainly observed during solar maximum periods. 

This work focuses on cosmic-ray protons and helium observations gathered above 70 MeV n$^{-1}$  with a particle detector (PD) placed on board the European Space Agency (ESA) LISA Pathfinder spacecraft \citep{lisapf1,lisapf2,armano2016,armano2017a} orbiting around the Lagrange point L1 at 1.5$\times$10$^6$ km from Earth  between February 2016 and July 2017. The years 2016-2017 belonged to the declining phase of the solar cycle 24, characterized by a positive epoch of the GSMF (defined by the solar magnetic field directed outward at the Sun north pole \citep{apj1,apj2}). 
%global solar magnetic field allowed for the comparison of the LISA Pathfinder cosmic-ray data above 70 MeV n$^{-1}$ with the AMS data gathered above 450 MeV n$^{-1}$, a magnetic spectrometer experiment placed on board the Space Station and neutron monitor data.
%allowed for disentangling the intensity of short-term variations with respect to long-term variations during individual Bartels Rotations. 
%In this work it is focused on the quantitative percent variations of the GCR flux above 70 MeV and above tens of GeV on the basis of the comparison of the LPF data with those from polar neutron monitors associated with individual high-speed solar wind streams with both a visual inspection and a neural network study. 
The role of isolated and interplaying IP structures  in modulating the cosmic-ray flux as a function of the energy is discussed here by comparing the LPF data with those  gathered during the same period of time above 450 MeV n$^{-1}$ by the AMS-02  magnetic spectrometer experiment hosted  on board the International Space Station (ISS; \cite{aguilar2018}) and with those from polar neutron monitors (\url{www.nmdb.eu}). It is observed that similar interplanetary structures   produce  different effects on cosmic rays. In  \cite{apj1,apj2,grim19} we carried out an attempt to interpolate the energy differential fluxes of the GCRs just before and at the dip of  Forbush decreases (FDs) \citep{forbush1,forbush2,forbush3,cane1}, sudden drops of the GCR intensity associated with the passage of ICMEs and shocks, with the aim  of providing  empirical relationships on the flux variation during the event evolution. A similar approach is adopted in this work for recurrent GCR flux variations.  
%Several lessons were learned with LISA Pathfinder meant for the technology testing of the Laser Interferometer Space Antenna (LISA), the  ESA mission  for low-frequency (10$^{-4}$ Hz - 10$^{-1}$Hz) gravitational wave detection in space  scheduled for launch in 2034 . In particular,  
The outcomes of this work and data from cosmic-ray and solar physics experiments will be used to study the LISA \citep{amaro2017} mission noise force induced by the charging process on the gold-platinum free-falling test masses \citep{grim15}. 
In particular, lessons learned with LPF about the relevant role of GCR short-term variations in generating low-frequency noise associated with  the test-mass charging process indicate that a continuous monitoring of the energy dependence of GCR short-term variations on board space interferometers will be mandatory \citep{armano2017b,mattia}. 
Energy, space and time short-term variations of the GCR flux will be monitored with particle detectors placed on board the LISA spacecraft  and neutron monitor observations.    
% of the Laser Interferometer Space Antenna (LISA), the  ESA mission meant for low-frequency (10$^{-4}$ Hz - 10$^{-1}$Hz) gravitational wave detection in space  scheduled for launch in 2034 \cite{amaro2017}. 
%LISA will consist of three satellites orbiting the Sun at 1 AU  50 million km behind the Earth.
%meant for low-frequency (10$^{-4}$ Hz-10$^{-1}$Hz) gravitational wave detection in space 
 Galactic cosmic-ray short-term variations will be also considered to estimate the varying background noise on the visible and ultraviolet images  of Metis, the coronagraph flown on board the ESA-NASA (National Aeronautics and Space Administration) Solar Orbiter launched from Cape Canaveral on February 9, 2020 \citep{andre,solwind14,metis19}. Finally, recurrent GCR flux depressions observed in L1 allow for monitoring HSS passage at the origin of  geomagnetic disturbances of weak to moderate intensity \citep{tsurutani}. 

\section{The particle detector on board LISA Pathfinder} \label{sec:style}

The LPF spacecraft  was launched from the Kourou base in  French Guiana on December 3, 2015 on board a Vega rocket.
The spacecraft  reached its operational six month orbit  around the Earth-Sun Lagrangian point 
L1 at 1.5 million km from Earth (in the Earth-Sun direction) at the end of January  2016.
The LPF elliptical orbit was inclined at about 45 degrees with respect to the ecliptic plane.
Orbit minor and major axes were approximately of 0.5 million km and 0.8 million km, respectively. The spacecraft spinned 
around its own axis  in approximately six months.  The LPF satellite carried two, nearly 2 kg cubic gold-platinum free-floating test masses that played the role of mirrors
of the interferometer.  Protons and nuclei with energies larger than 100 MeV n$^{-1}$  penetrated or interacted in   13.8 g cm$^{-2}$ of spacecraft and instrument material thus charging the LPF test masses.
The test-mass charging  due to galactic and solar energetic particles (SEPs)
results in   noise  force limiting the sensitivity of space interferometers mainly below 1 mHz \citep{armano2017b}. As a result, a PD \citep{can11} was placed 
on board LPF 
to monitor the overall incident GCR and solar particle fluxes  \citep{sha06}. The PD was mounted on a honeycomb  satellite wall at about one meter distance from test masses and surrounding instrumentation in order to limit the impact of secondary particles in  affecting the response of the instrument \citep{henrique,daniel}. The PD viewing axis was oriented 
along the Sun-Earth direction. The detector consisted of
two $\sim$ 300 $\mu$m thick silicon wafers
%300 $\mu$m thick,                                                                                                 
of  1.40 x 1.05 cm$^2$ area each,   separated by 2 cm  and placed in
a telescopic arrangement. For particles with  energies $>$ 100 MeV n$^{-1}$ and an isotropic incidence,
the instrument geometrical factor  was  found to be energy independent and equal to  9 cm$^2$ sr  for each silicon layer.  The total geometrical factor for particle counting was 17 cm$^2$ sr. In coincidence mode (particles traversing both silicon wafers), the geometrical factor was  0.9 cm$^2$ sr. The silicon wafers were placed inside a shielding copper  box of 6.4 mm thickness meant to stop particles with energies below 70 MeV
n$^{-1}$. The PD  allowed for the counting of
 protons and helium nuclei
 traversing  each silicon layer and for the measurement of ionization energy losses of particles in coincidence mode.
%The PD  data are stored  in the form of histograms over periods of 600
%seconds and then sent to the on-board computer. 
The maximum allowed detector  counting rate
was 6500 counts s$^{-1}$ on both silicon wafers, corresponding to an event integrated  proton fluence of   10$^8$ 
protons cm$^{-2}$ at energies $>$ 100 MeV.
%An increasing counting rate starting from  more than 120 counts per sampling time of 15 seconds was observed over\
% the mission lifetime due to a decreasing solar activity intensity.                                                
In coincidence mode  up to 5000 energy deposits per second could be stored on the on board computer.  Particle counts only were used for the analysis presented in this work (the corresponding instrument GF was 17 cm$^2$ sr). Particle ionization energy losses would have been used for SEP evolution monitoring. No SEP events  were observed during the LPF mission operations with associated  proton fluences above a few tens of MeV n$^{-1}$  overcoming that of GCRs. 
%The Nymmik model 
%\citep{nymmik1,nymmik2} 
%allowed for the estimate of the rate of occurrence of SEP events with flu\
%ences larger than the saturation limit
%for the period the LPF spacecraft was orbiting around L1.
%The  occurrence  of SEP events with fluence
%$>$ 10$^8$ protons cm$^{-2}$  at energies $>$ 30 MeV, for instance,  was estimated to be  less
%than 1 per year  for the expected solar activity level during the solar cycle 24 
%\citep{sto08} 
%(aggiungi referenza di Grimani et al. CQG 29 2012).

\section{Galactic cosmic-ray flux variations with time}
%No SEP events  were observed during the LPF mission operation with associated  proton fluences
%above a few tens of MeV n$^{-1}$  overcoming that of GCRs.
% overcoming that of protons of galactic origin.
%above a few tens of MeV n$^{-1}$  overcoming that of protons of galactic origin, 
A continuous monitoring   of long and short-term variations of cosmic rays  of galactic origin  was allowed  during the LPF mission elapsed time. 
%Due to a decreasing solar activity during the mission about the cosmic-ray counting rate varied from 120 counts s$^{-1}$ through over  150 counts s$^{-1}$. The percent variations of hourly avaraged data were contained below 1\% over the entire mission lifetime. 

%A particle detector flown on board LISA Pathfinder allowed to gather galactic cosmic-ray proton and helium nucleus fluxes between 2016 and 2017 during the decreasing phase of the solar cycle 25. The number of observed sunspots during the data taking period is reported in Table 1. Previous work on GCR flux short-term variations with LISA Pathfinder  have been reported in (refs). In this previous work we have studied   

\subsection{Long-term variations}

Long-term variations of the GCR flux are those characterized by  durations $>$ 1 year. The GCR flux modulation associated with the eleven year solar cycle and the twenty-two year GSMF polarity change are widely studied in the literature both on Earth with neutron monitors (NMs; see for instance \cite{storini}) and in space with balloon  \citep{Shikaze2007154} and satellite experiments \citep{potgieter}. The monthly averaged sunspot number (SSN; \url{www.sidc.be/silso/datafiles\_old}) is the most widely used  proxy to monitor the variations of the solar activity correlated with the solar modulation parameter within the symmetric model of GCR energy spectra in the force-field  approximation  by Gleeson and Axford (G\&A; \cite{glax68}). This model  allows to estimate  the cosmic-ray intensity in the inner heliosphere by assuming an interstellar energy spectrum and
 a solar modulation parameter that basically represents the energy loss of cosmic rays while propagate from  the interstellar medium to  the point of observations. The G\&A model is found to reproduce the  GCR measurements at 1 AU in the energy range  from tens of MeV to hundreds of GeV during GSMF positive polarity epochs \citep{grim07}. It is worthwhile to point out that different values of the solar modulation parameter are set when data are compared to model predictions if different GCR energy spectra at the interstellar medium are considered. Voyager 1 measured the interstellar spectra of ions and electrons below 1 GeV \citep{stone2013}, however, the solar modulation parameter reported in \url{http://cosmicrays.oulu.fi/phi/Phi\_mon.txt}, was estimated according to the \cite{burger2000} interstellar proton spectrum \citep{uso11}. Therefore, solar modulation parameter and interstellar particle spectra must be properly adopted.
%In case  predictions of   GCR flux in future years are carried out when the solar modulation and the solar polarity are known, this approach can be reasonably assumed, conversely if GCRs are propagated from interstellar medium through the inner heliosphere at a certain time the interstellar measurements observed by the Voyager 1 should be considered (ref). 
The  G\&A  model takes into account particle diffusion and convection processes, however during negative (positive) polarity periods of the GSMF, positive (negative) GCR particles  undergo also a global drift motion from the solar equator (poles) towards  the poles (equator). Positive (negative) particle fluxes are more modulated during negative (positive) polarity periods and a larger solar modulation is estimated with respect to an analogous solar modulation during a positive polarity epoch \citep{grim07}.
% but this intensity the modulation is different negative solar cycles.   
%   while  the particle drift process must be also considered in addition to diffusion and convection  during negative polarity periods of the GSMF. 
%Voyager 1 measured the IS spectrum of protons but the solar modulation parameter database by Usoskin et al. (ref) was estimated according to the Burger et al.; xxx IS spectrum 
As a result, the monthly solar modulation parameter reported in \url{http://cosmicrays.oulu.fi/phi/Phi\_mon.txt}  
 is representative of the  overall effects of long and short  GCR flux variations \citep{uso17}. This empirical approach allows us to make  average predictions of the  GCR flux when the solar activity   is  known (see  \cite{apj2}). However, to study the performance of very sensitive instruments, for instance, the role of short term variations must be considered separately from the long term ones.\\
Observations carried out with the BESS  \citep{Shikaze2007154} magnetic spectrometer  experiment flown on  balloons several times during more than one solar cycle after 1997  allowed to investigate the energy dependence of the solar modulation and solar polarity inferred from proton and helium  differential fluxes. 
%Many other previous short flights of balloon-borne experiments, despite  carrying also magnetic spectrometers, did not allow to monitor the long-term effects of the solar modulation and solar polarity (see for instance paper coi dati di MASS89).  
More recently, the  experiments AMS-01 on the Space Shuttle Discovery \citep{ams01},   PAMELA on a satellite \citep{adriani} and   AMS-02 on the International Space Station (ISS) \citep{aguilar2018} provided  continuous cosmic-ray data monitoring in space in the last 15 years. 
%All the cited experiments mentioned above carried magnetic spectrometers for GCR flux differential flux measurements. 
PAMELA was launched in 2006 and remained into a quasi-polar elliptical orbit around the Earth  between the end of the solar cycle 23  and the solar cycle 24, mostly during a negative polarity period. AMS-02 was installed in 2011 on the ISS and gathered data for the majority of time during  a positive polarity epoch after the last polarity change from - to + at the end of 2013. PAMELA observations extended down to 70 MeV n$^{-1}$ and allowed for monitoring the most populated region of the GCR energy spectrum  which is a fundamental requirement for monitoring short-term GCR flux variations, while AMS-02 provided data above 450 MeV n$^{-1}$ due to the shielding effect of the geomagnetic field along the ISS orbit. 
%All the cited experiments mentioned above carried magnetic spectrometers for GCR flux differential flux measurements.
%The limit of this experiment is that being placed on the Space Station, only data above 450 MeV n$^{-1}$ are available.
%Many other previous short flights of balloon experiments despite these last ones were carrying magnetic spectrometers did not allow to monitor the long-term effects of the solar modulation and solar polarity (see for instance).  
%To our knowledge, none of the aforementioned experiments carried out observations of the GCR flux during individual recurrent short-term variations.
PAMELA GCR short-term variation observations were reported for instance in \citet{pamFD, pamAPJ, pamAPJL, pamICRC}. Unfortunately, it was not feasible for us to infer from paper figures the energy dependence of individual GCR flux short-term variations on time scales of hours or days at the most.

A decreasing solar activity was observed during the LPF mission elapsed time when the monthly averaged sunspot number 
varied from 56  in February 2016  through 18 in July 2017 \citep{apj1}. 
 During this same period the solar modulation parameter and the LPF cosmic-ray PD counting rate varied from 468 MV/c through 383 MV/c (\url{http://cosmicrays.oulu.fi/phi/Phi\_mon.txt}) and from about 120 counts
 through over  150 counts  every 15 seconds, respectively, revealing an average increase of 20\% of the GCR  proton and helium flux intensity   above 70 MeV n$^{-1}$ \citep{apj1,grim_sohe3}.
%, respectively the  The cosmic-ray counting rate varied from 120 counts 
%s$^{-1}$ through over  150 counts s$^{-1}$ every 15 seconds. 
%The percent variation of LPF hourly avaraged cosmic-ray data estimated with respect to the average value observed during each Bartels rotation (BR)   presented statistical uncertainties $<$ than 1\% over the entire mission lifetime. It is recalled here that the BR number corresponds to the number of 27-day rotations of the Sun since February 8th, 1832.

%For the study of the energy dependence of short-term recurrent variations the LISA Pathfinder data gathered above 70 MeV n$^{-1}$ are compared to observations gathered on neutron monitors placed on Earth at different latitudes allowing for the percent GCR variation monitoring above 10 GeV. A similar approach was applied to Forbush decreases (non-recurrent GCR depressions) presented in Armano et al. 2019.  

\subsection{Short-term variations}

GCR flux short-term variations are characterized by durations $<$ 1 month and result associated with the transit of HSS and IP structures. The GCR flux short-term variations can be either recurrent and non-recurrent. 

The first kind is mainly associated with CIRs generated by HSS originating from coronal holes and overtaking leading slow solar plasma \citep{richardson1}. Recurrent GCR flux short-term variations  present the same periodicities of the Sun rotation period and higher harmonics (see for instance \cite{apj1}).

Intense non-recurrent GCR flux drops  associated with ICME and shock transit are known as FDs. These short-term GCR flux depressions show up to 25\% intensities in Earth neutron monitors  \citep{simoapj}.  
%The percent variations of LPF hourly avaraged cosmic-ray data estimated with respect to the mean value observed during each Bartels rotation (BR)   
%presented statistical uncertainties $<$ than 1\% over the entire mission lifetime thus allowing for the monitoring of short-term periodicities. It is recalled here that the BR number corresponds to the number of 27-day rotations of the Sun since February 8, 1832
%(Armano et al.; 2018 and Benella et al.; 2018, 2019).  
%The LPF data revealed that above 70 MeV n$^{-1}$ the monthly GCR flux varied at most by 15\%  in 2016-2017. This GCR flux intensity excursion is equivalent to that associated with a solar modulation parameter increase of about 100 MV/c. 
%Therefore, the role of GCR short-term variations  must be correctly taken into account to carry out precise estimates of instruments performance in space. 

GCR recurrent and non recurrent short-term variations have been widely studied in the last sixty years with Earth neutron monitors  \citep{forbush2,forbush3,cane1}. However, in \cite{apj1,apj2} it was pointed out that the whole evolution of weak Forbush decreases
%sudden drops of GCR flux intensities associated with the propagation of the IP counterparts of coronal mass ejections and shocks, 
and recurrent variations showing intensities $<$ 10\% in space  
cannot be studied with neutron monitors measurements that appear modulated by $<$ 3\%. 
As a matter of fact, neutron monitor observations  are sensitive to secondary particles generated by primaries incident at the top of the atmosphere with energies above 500 MeV but vary of the same amount of the GCR integral flux, thus providing a direct measurement of the same, at energies above {\it effective energies} 
%\citep{gil3} 
(which range between  11-12 GeV and  above 20 GeV for near-polar 
and equatorial stations, respectively; see \cite{gil17}). Noteworthy, a large sample of data below 100 MeV/n  
will be available  soon with the  EPD (Energetic Particle Detector;  \cite{epd}) on board Solar Orbiter for instance. Unfortunately, the small geometrical factor in the cosmic-ray energy range limits the capability of this instrument to follow the dynamics of the GCR flux short-term variations smaller than one day. 

%The  data gathered by the LISA Pathfinder mission in the years 2016-2017 during individual and Bartels rotations allowed to disantangle the role of long and short-term variations of the GCR flux during those years, after comparison of these data with those punlished by AMS-02 during the same Bartels rotations. 
%By comparing the LISA Pathfinder data to those gathered during the same period by the  AMS-02 observations, 
%It was observed that short-term variations, affect GCR model predictions that take into account the long-term solar modulation effect only. 

\section{GCR flux recurrent depressions and interplanetary plasma parameters during the LISA Pathfinder elapsed time}
 The LPF hourly averaged proton and helium integral flux measurements above 70 MeV n$^{-1}$  were characterized by statistical uncertainties of 1\% over the entire mission duration. In order to limit the effect of the solar modulation decrease in 2016-2017, the percent variations of these  data were calculated with respect to the average value observed during each Bartels rotation (BR) through the whole period of the mission operations (see \cite{apj1}).
%The LPF data revealed that above 70 MeV n$^{-1}$ the monthly GCR flux varied at most by 15\%  during the Bartels rotation (BR) 2496 (ref). 
It is recalled 
here that the BR number corresponds to the number of 27-day rotations of the Sun since February 8, 1832
%Detailed studies of the periodicities present in the whole LPF data set were reported in                        
\citep{apj1,benny19b,benny19a}. Data gathered during the BRs 2493, 2494, 2500 and 2503 are reported in the following.
%It is worthwhile to point out that  during periods characterized by the presence of equatorial coronal holes or equatorward extensions of polar coronal hole                                                                                                      
The LPF data revealed that above 70 MeV n$^{-1}$ the monthly GCR flux varied up to 15\%  during the BR 2496 \citep{apj1,grim19}. A similar overall GCR flux decrease is
observed  when the solar modulation parameter increases by about 100 MV/c. As a result,  it is important to disentangle the role of long and short-term cosmic-ray variations.
Forty-five GCR flux recurrent variations were observed during the LPF mission elapsed time from  March 2016 (BR 2491) through July 2017 (BR 2509). %The average characteristics of recurrent  GCR flux depressions observed  in space with LPF are reported in table 1.  
In Fig. 1  the LPF GCR percent flux variations during the BR 2493 (April 27, 2016 - May 23, 2016) are compared to contemporaneous IP plasma parameter and magnetic field data. 
The association between  GCR flux drops and solar wind speed increase at the passage of HSS is observed all through the data sample (see  \cite{richardson} for instance for a previous study). The following  criteria for selecting the GCR recurrent depressions observed on board the LPF spacecraft  were applied: 1) depression commencements were set at the beginning of each continuous
decrease of the GCR flux observed
 for more than 12 hours; 2) the whole duration was $>$ 1 day and 3) the amplitude was $>$ 1.5\%. In general, the peaks of the GCR flux  are observed when the solar wind speed presents values smaller than 400 km s$^{-1}$ and the IP  magnetic field remains below 10 nT \citep{apj1,apj2,grim19}.  
%GCR flux periodicities common also to  BV or BV$^2$ timeseries, where B is the interplanetary magnetic field intensity and V the solar wind speed were discussed by (refs).  
%Periodicities associated with the Sun rotation period and higher harmonics are detected in cosmic rays both on Earth and in space, in the solar wind parameters and interplanetary magnetic field. However, because of the energy dependence of each process, characterized by enhanced by the GCR flux modulation observations. ,  Earth observations do not allow to estimate the GCR intensity variations in space. 
However, even when the above IP parameters are higher than the indicated values and the GCR flux appears modulated, the amplitude of the modulation appears different depending on the GCR flux intensity before the perturbation and on the interplay 
of subsequent HSS.
For instance, the decrease of the GCR flux between May 7  and May 9, 2016 in Fig. 1 was  of 5\% due to a solar wind increase of about 300 km s$^{-1}$ from about 400 km s$^{-1}$ through 700 km s$^{-1}$. A double decrease is observed between May 16 and May 21, 2016 with a solar wind increase of less than 200  km s$^{-1}$ above 400 km s$^{-1}$. The difference is ascribable to the pre-HSS passage intensity of the GCRs: compatible with the average value observed during the whole BR 2493 in the first case and equal to the maximum value observed during the same BR in the second case.  Maximum and minimum variations with respect to monthly averages observed during the mission operations were +9\% and -7\% observed during the BR2495 and BR2502, respectively.  
The role of interplaying structures can be observed  by comparing  Figs. 1 and 2 where the LPF GCR data during the BRs 2493 and 2494 are shown.
The last week of these two BRs is characterized by a GCR flux depression beginning on May 16-17 during the BR 2493 and 
June 12-13 during the BR 2494. The decrease phase is associated with the transit of  recurrent HSS.  During the BR 2493 the GCR flux is observed to recover after one-day $plateau$  \citep{apj1} because  
the  speed of the solar wind goes down to  $\simeq$ 400 km s$^{-1}$ while during the BR 2494 the GCR flux recovery phase is  disturbed  for 6 days by the  passage of several superposing HSS with the solar wind speed remaining well above 400 km s$^{-1}$. 
%In particular, it is interisting to notice that the depressions of cosmic rays are regulated by the interplay of subsequent IP structures.                                                                                                                                    
The  energy dependence of the GCR flux variations is inferred from  Figs. 3 and 4 and Table 1  from the comparison of  GCR flux decreases observed with LPF during the BR 2493 and BR 2494 above 70 MeV n$^{-1}$ to contemporaneous 
hourly averaged polar neutron monitor measurements representing the GCR flux variations above the effective energy of 11 GeV
(see \cite{apj1} and references therein for details).  
%The following  criteria for selecting the GCR recurrent depressions observed on board the LPF spacecraft  were applied: 1) depression commencements were set at the beginning of each continuous 
%decrease of the GCR flux observed
% for more than 12 hours; 2) the whole duration was $>$ 1 day and 3) the amplitude was $>$ 1.5\%.

%Above 70 MeV the proton dominated LPF measurements in space  present 
%intensities of about 10\%  while contemporaneous  polar NMs observations representative of the GCR flux above 11-12 GeV, vary by  $\pm$2\%.

\begin{figure}
  \begin{center}
  \plotone{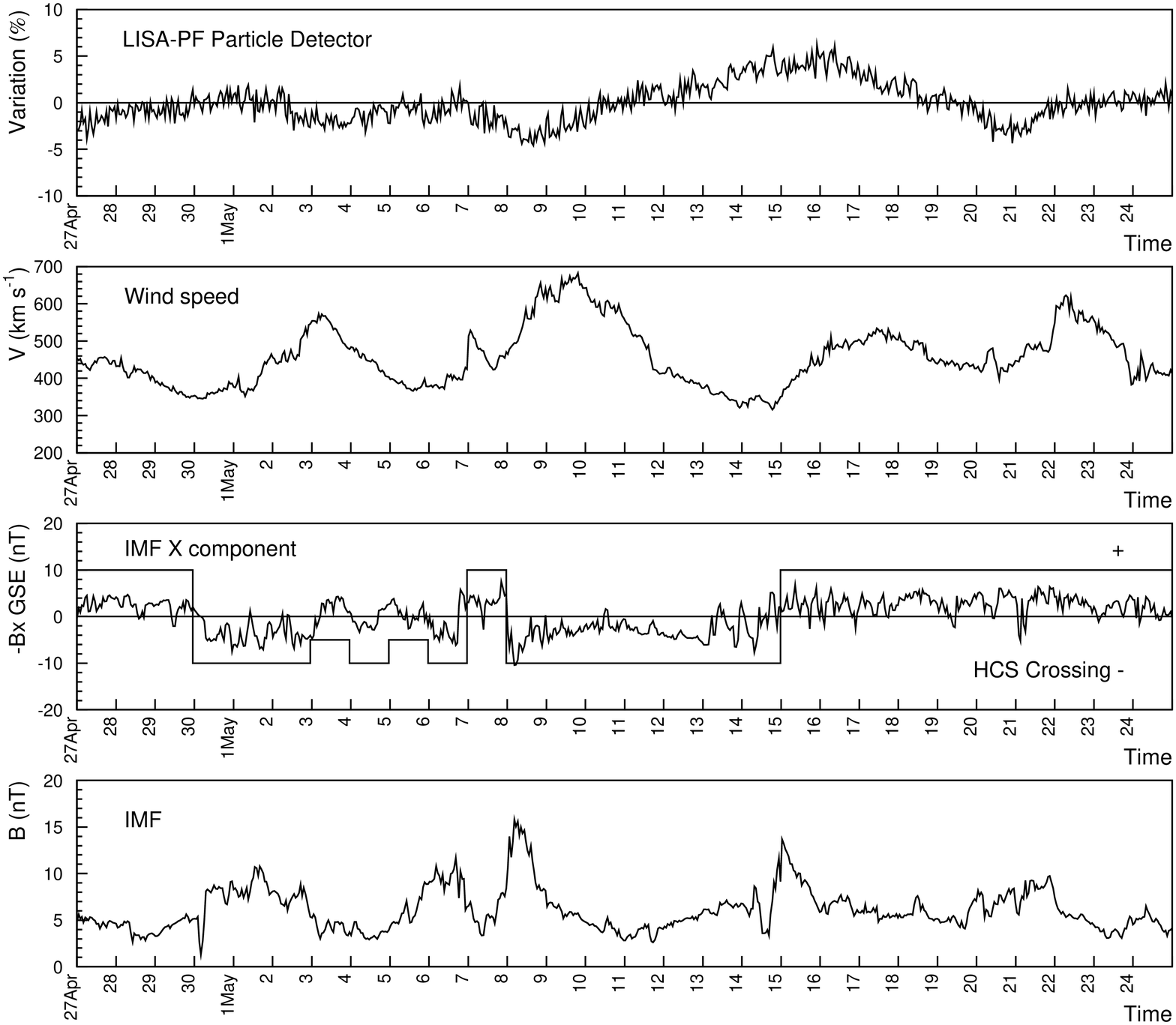}
  \caption{LPF PD counting rate percent variations during the BR 2493 (April 27, 2016 - March 23, 2016; first panel).  Solar wind speed 
(second panel), IMF radial component (third panel) and IMF intensity (fourth panel) contemporaneous measurements, gathered by the ACE 
experiment, are also  shown. HCS crossing (\url{http://omniweb.sci.gsfc.nasa.gov./html/polarity/polarity\_tab.html}) is indicated in the 
third panel.
%Periods of time during which the solar wind speed (V) and  the magnetic field (B) intensity remain below and above 400 km s$^{-1}$ and  10 nT, respectively, are shown in the second and fourth panels.  Decrease, plateau and recovery periods of each GCR depression are represented by red, blue and cyan lines in the first panel.                                                                                 
      }
  \label{figure1}
 \end{center}
\end{figure}

\begin{figure}
  \begin{center}
  \plotone{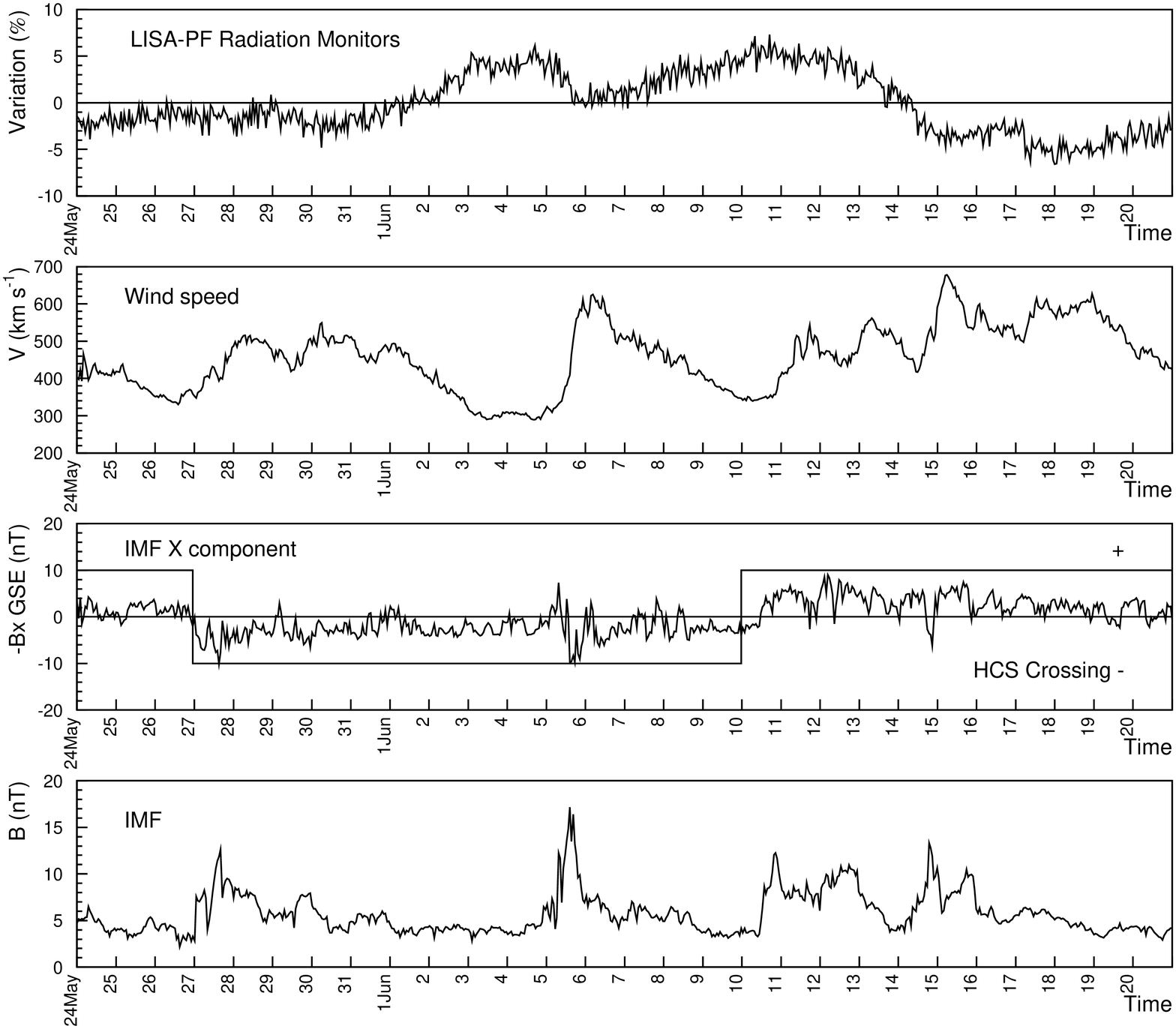}
  \caption{Same as Fig. 1 for the BR 2494 (May 24, 2016- June 20, 2016).}
  \label{figure1}
 \end{center}
\end{figure}

\begin{figure}
\plotone{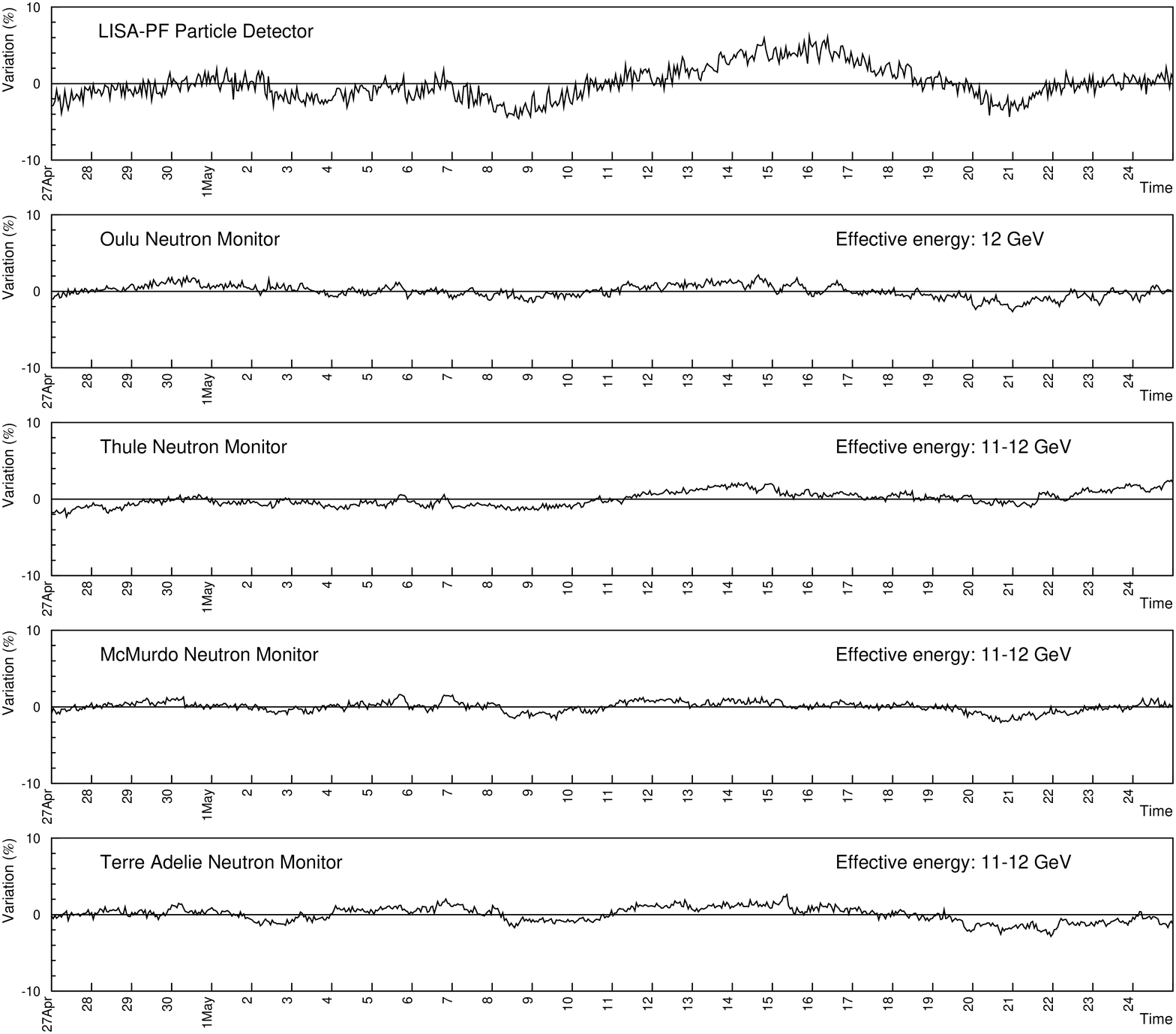}
  \caption{Comparison of hourly averaged LPF PD counting rate percent variations with contemporaneous, analogous measurements 
of polar neutron monitors  during the BR 2493
(April 27, 2016 - May 23, 2016).}
  \label{figure1}
 \end{figure}

\begin{figure}
  \plotone{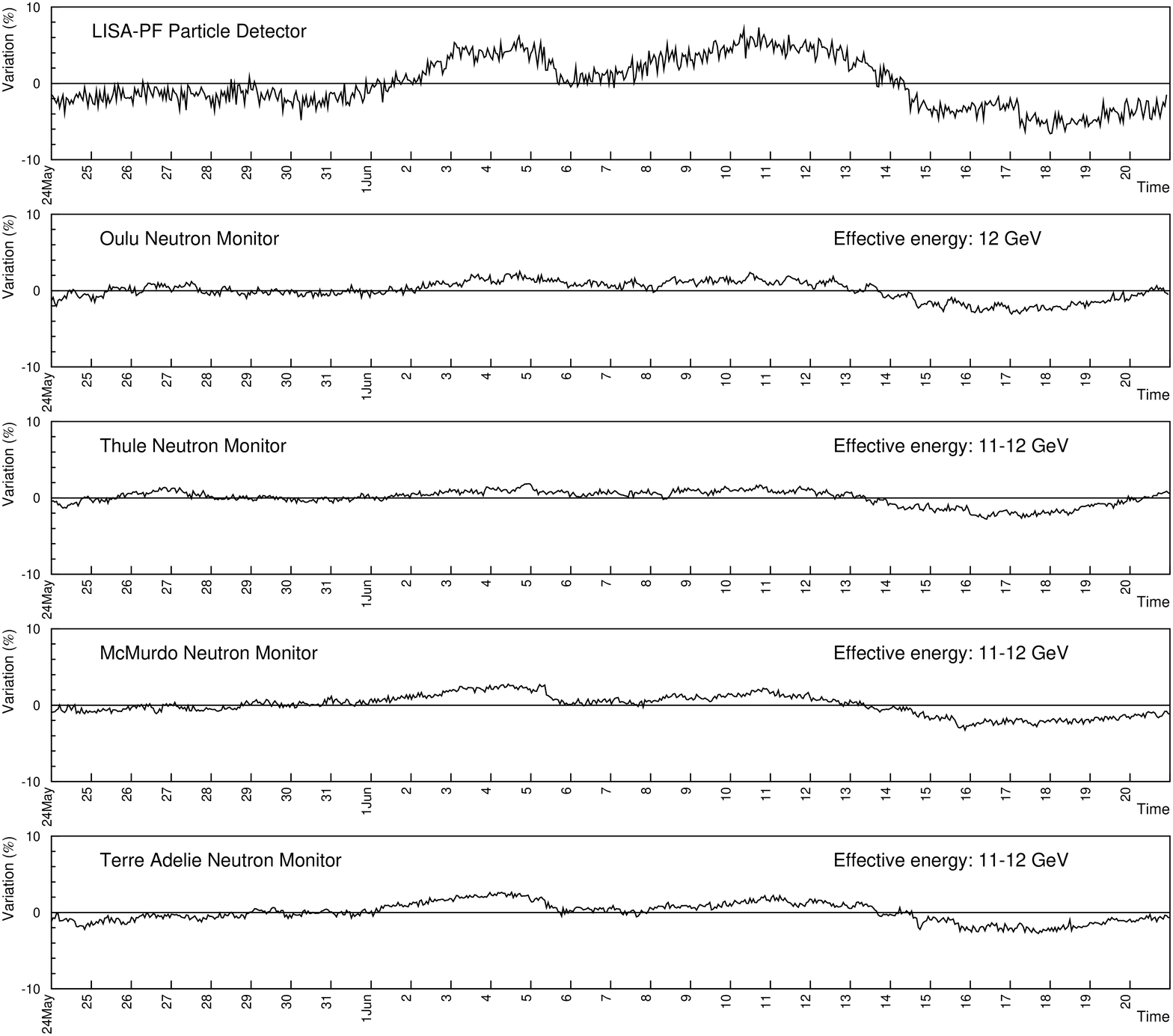}
  \caption{Same as Fig. 3 for the BR 2494 (May 24, 2016- Jyne 20, 2016).}
  \label{figure1}
\end{figure}

%\startlongtable
\begin{deluxetable}{ccccccc}
\tablecaption{Energy dependence of the GCR integral flux percent change  (PC) at the maximum of the six recurrent depressions observed on board LPF during the BRs 2493-2494 (see Figs. 3 and 4) above 70 MeV n$^{-1}$ and with  NMs above effective energies: 11-12 GeV for  polar stations; 12 GeV for Oulu NM. Onset and dips of each depression were set according to the LPF data. Percentages were rounded to 0.5 percent. \label{tab:table}}
\tablehead{
  \colhead{LPF depression onset} &\colhead{LPF  depression dip} & LPF &\colhead{Terre Adelie}& \colhead{McMurdo}& \colhead{Thule}&\colhead{Oulu}\\
\colhead{Time} & \colhead{Time} & \colhead{$>$ 70 MeV} &\colhead{$>$ 11 GeV}  &\colhead{$>$ 11 GeV} & \colhead{$>$ 11 GeV}& \colhead{$>$ 12 GeV}\\
\colhead{} & \colhead{} & \colhead{PC} &\colhead{PC} & \colhead{PC} & \colhead{PC}&  \colhead{PC}
}
\startdata
     2016 May 2 02.40 UT & 2016 May 3 10.35 UT& 3\% & $<$1\% & $<$1\% & $<$1\% & $<$1\%\\
     2016 May 6 18.45 UT & 2016 May 8 13.20 UT& 4.5\% & 2\% & 2\% & 1\% & 1\%\\
     2016 May 16 08.00 UT & 2016 May 20 20.15 UT& 8\%  & 3\% & 2\% & 2\% & 2\%\\
     2016 May 29 01.00 UT & 2016 May 30 10.50 UT& 3\% & $<$1\% & $<$1\% & $<$1\% & $<$1\%\\
     2016 June 5 04.15 UT & 2016 June 6 02.30 UT& 4.5\% &2.5\% & 2.5\% & 1.5\% & 1.5\%\\
     2017 June 12 13.15 UT & 2016 June 17 22.25 UT& 9.5\%  & 3.5\% & 3.5\% & 3.5\% & 3.5\%\\
\enddata
\end{deluxetable}

\begin{figure}[ht]
\plotone{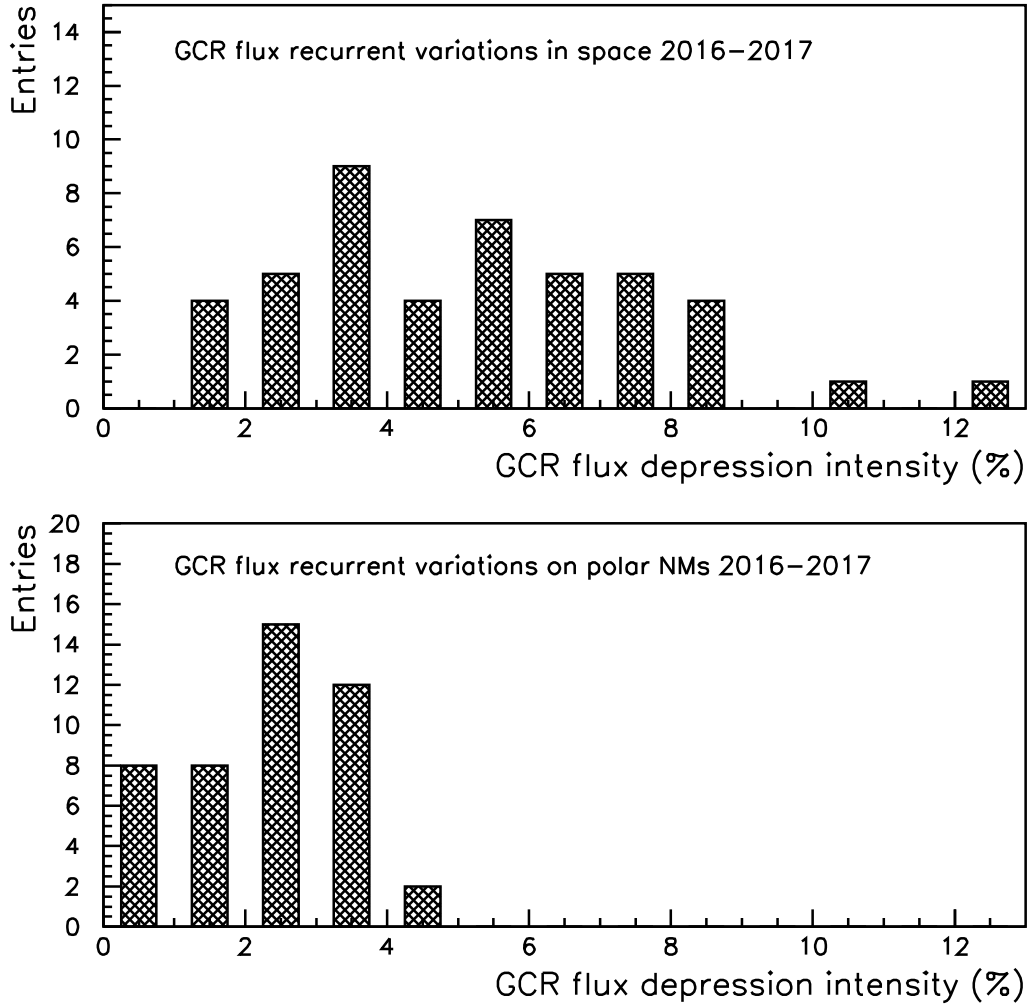}
\caption{\label{gcr14} Comparison of GCR flux recurrent percent decreases observed in space on board LPF in
2016-2017 above 70 MeV n$^{-1}$ (top panel) and on average on  Earth with polar neutron monitors above 11 GeV n$^{-1}$ (bottom panel).}
\label{figure1}
\end{figure}

\begin{figure}[ht]
\plotone{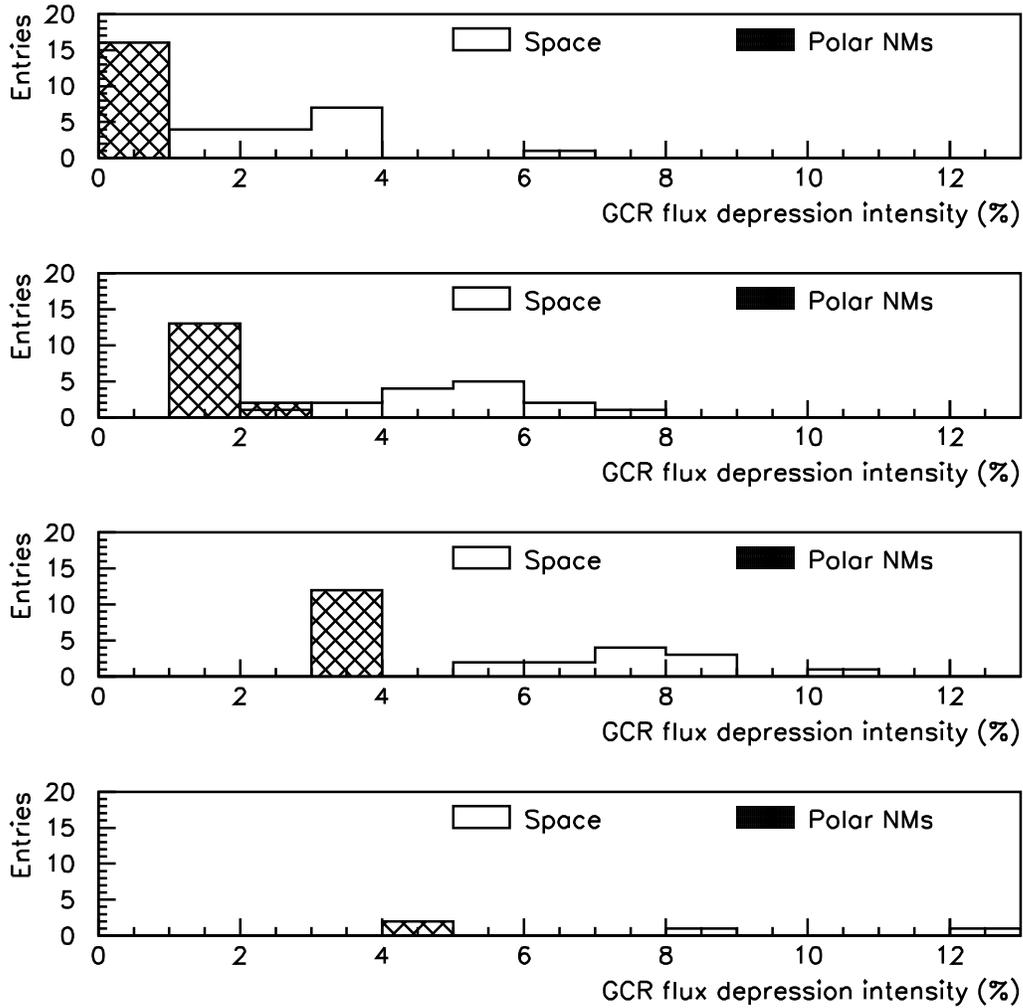}
  \caption{\label{gcr14} Comparison of GCR flux percent recurrent depressions observed on Earth polar neutron monitors
(hatched area) in the intervals (0-1\% - 1-3\% - 3-4\% 4-5\% from top to bottom panels) with corresponding variations observed in
 space on board LPF in 2016-2017 (white area).}
\label{figure1}
\end{figure}

\begin{figure}[ht]
\plotone{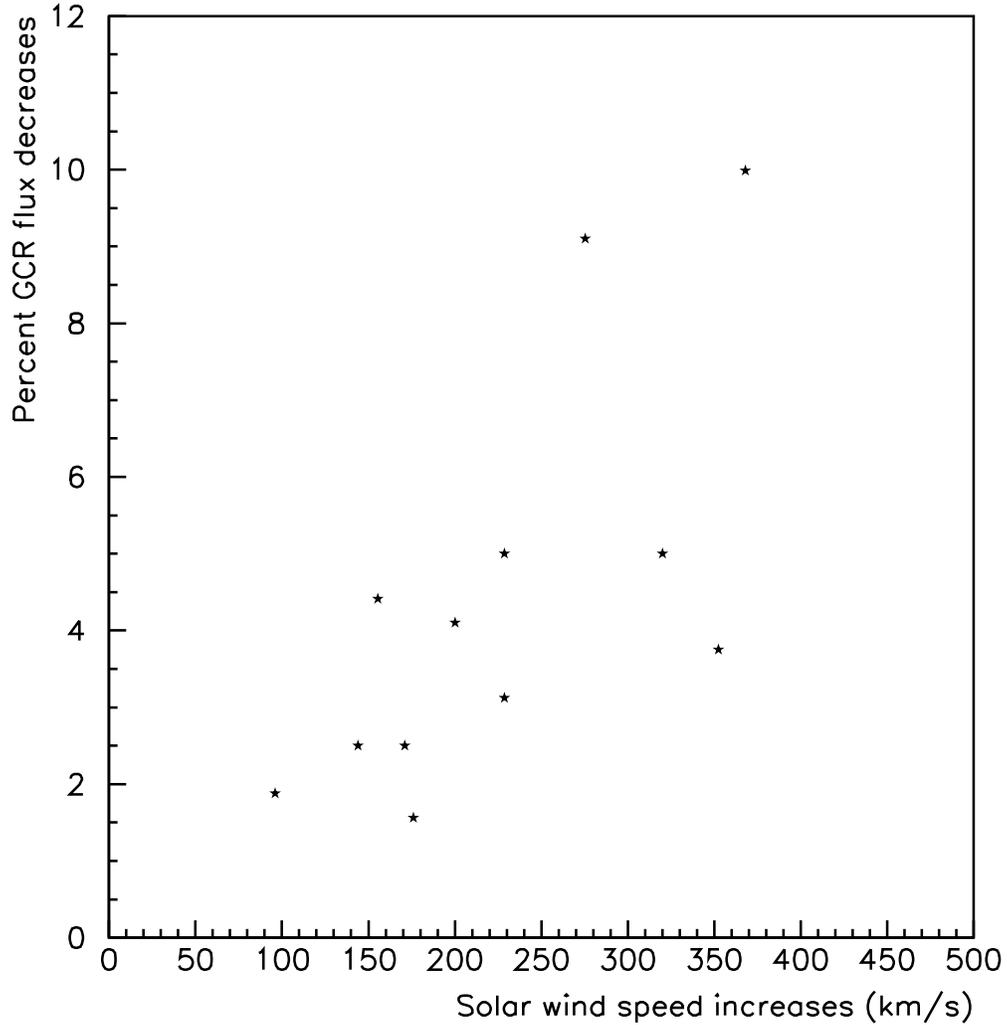}
\caption{\label{gcr14} Solar wind speed increases   and corresponding
GCR flux percent decreases  at the passage of HSSs in 2016-2017.}
\label{figure1}
\end{figure}

\begin{figure}[ht]
\plotone{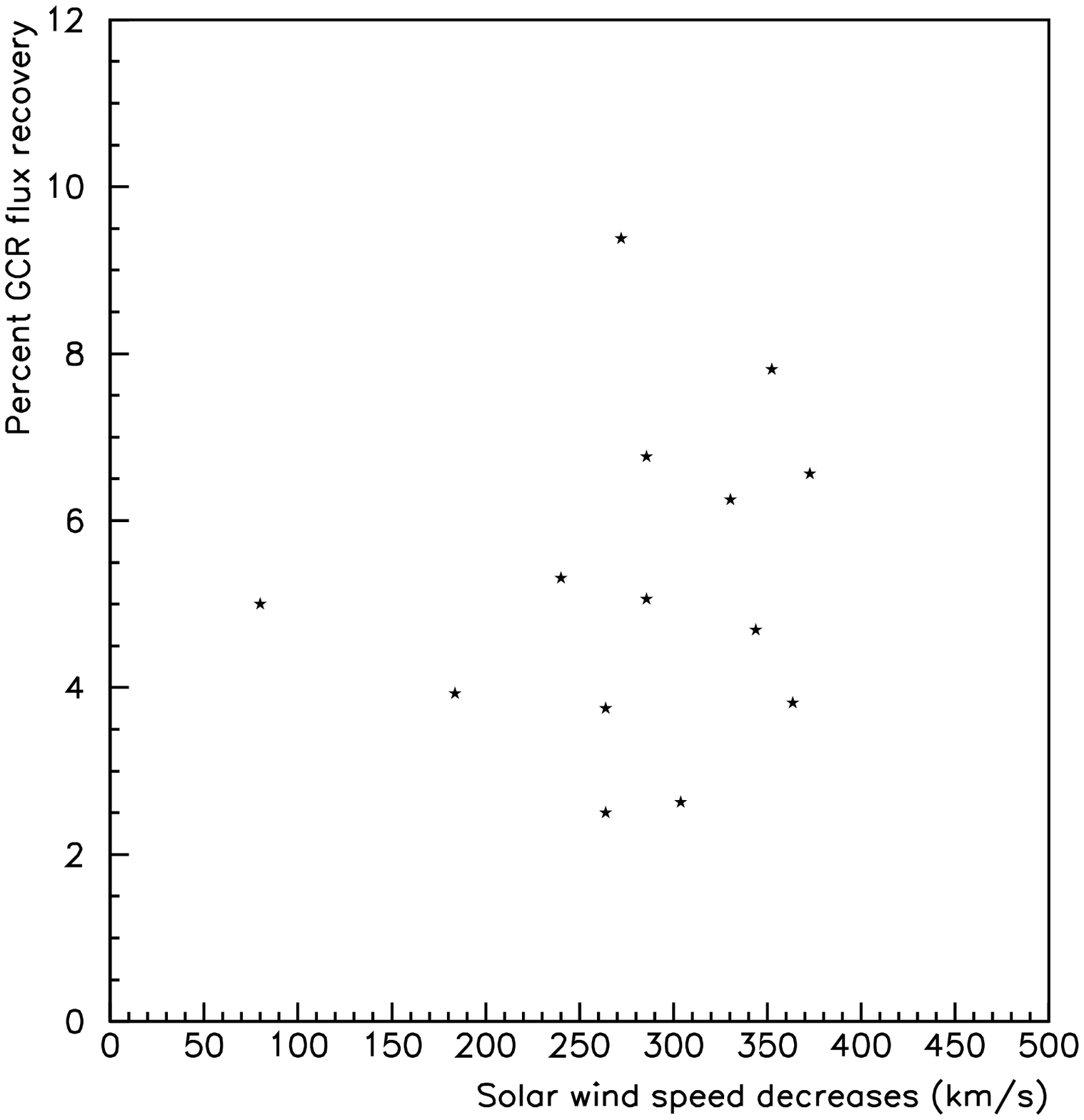}
  \caption{\label{gcr14}  Solar wind decreases   and corresponding
GCR flux percent recoveries  in 2016-2017.}
\label{figure1}
\end{figure}

%In Fig. 2 the LPF  GCR flux data gathered during the same BR are compared to contemporaneous hourly averaged polar neutron monitor observations. The energy dependence of the GCR flux variations is inferred from the comparison of these data. Above 70 MeV the proton dominated LPF measurements in space show more marked variations than those  gathered with  neutron monitors which measurements are  representative of the GCR flux above 11-12 GeV. 
%The  criteria for selecting GCR depressions were the following: depression commencements were set at the beginning of each continuous decrease of the GCR flux observed
 %for more than 12 hours, the whole duration was $>$ 1 day and the amplitude $>$ 1.5\%.
%, therefore neutron monitors do not allow for studying the effects of interplanetary structures in space in depleting the low energy part of the GCR spectrum.  
The average characteristics of recurrent  GCR flux depressions observed  in space with LPF and contemporaneously with  polar neutron monitors are reported in Table 2. 
%Despite the average GCR flux drops  gathered in space and on Earth agree within 1 $\sigma$,  
Neutron monitor data present a peaked  distribution, while 
data in space show a  flat distribution as it can be observed in 
\begin{table}
\centering
\caption{\label{table1} Average characteristics of GCR flux recurrent short-term depressions observed with LPF in space in 2016-2017. A comparison
is carried out with contemporaneous GCR flux depression intensity observed with Earth  polar NMs. GCR flux short-term variations
gathered in space in 2016-2017 varied from 1.6\% through 12.3\% while on neutron monitors ranged from 1\% to 4.1\%.  The average duration of the depressions observed with polar neutron monitors was not estimated due to the difficulty to establish the onset and end times of each GCR flux decrease.}
\begin{tabular}{lll}%{@{}*{9}{l}}                                                                                              
\hline
\hline
 & Duration & Intensity \\
 & (Days)   & (\%)  \\
\hline
Space    &  9.2 $\pm$ 5.0 & \\
Space   &   & 5.1 $\pm$ 2.5 \\
Polar NMs & ----------- & 1.8 $\pm$1.6 \\
\hline
\end{tabular}
\end{table}
 Figs. 5 and 6  where the whole sample of the GCR flux recurrent  variations  observed in 2016-2017 with LPF  are reported. 
In particular, it can be observed that  while polar neutron monitors indicate  maximum variations of approximately 4\%,  the same variations in space above 70 MeV $n^{-1}$ are as large as more than 12\%.
% As a matter of fact, it is mandatory to fly radiation monitors on board interferometers for gravitational wave detection in space to estimate the noise on metal free-falling test-mass charging. 
The IP disturbances estimated to be at the origin  of each one of these variations are reported in \cite{apj1}. 
In 87\% of cases HSS transit only was associated with  recurrent GCR flux variations while the remaining 13\% of cases HCSC and ICME transit contributed in modulating  the GCR flux.  
%With respect to recurrent GCR flux variations the anticorrelation between solar wind velocity increase and cosmic-ray intensity were presented in (Refs Kondoh et al.; 1999, Armano et al.; 2018, 2019 Grimani et al. 2019). 
%Therefore during periods characterized by the presence of equatorial coronal holes or eqautorward extensions of polar coronal holes, the role of short-term variations must be taken into account for carrying out precise estimates of instruments performance in space. GCR short-term variations were widely studied with Earth neutron monitors. 
%Periodicities associated with the Sun rotation period and higher harmonics are detected in cosmic rays both on Earth and in space, in the solar wind parameters and interplanetary magnetic field. However, because of the energy dependence of each process, characterized by enhanced by the GCR flux modulation observations. ,  Earth observations do not allow to estimate the GCR intensity variations in space. In Armano et al.; 2018 the particle detector (PD) observations on board LISA Pathfinder  
%gathered in space between 2016 and 2017 allowed for the estimate of the proton and helium integral flux variations with 1\% uncertainty on hourly averaged 15-second data revealing the characteristics 
%of the profiles of GCR both recurrent and non-recurrent short-term variations above 70 MeV n$^{-1}$ during subsequent BRs are reported in figs 1-4. 
%Characteristics of interplanetary processes generating non-recurrent variations were reported in  Armano et al.; 2019. 

\section{Energy dependence of recurrent short-term GCR flux variations with LISA Pathfinder}
%The  criteria for selecting GCR depressions were the following: depression commencements were set at the beginning of each continuous decrease of the GCR flux observed for more than 12 hours, the whole duration was $>$ 1 day and the amplitude $>$ 1.5\%.
 Fourteen isolated leading edges and twelve trailing edges  of HSS modulating the GCR flux were studied. 
HSS were considered isolated if their passage was separated by at least one day from other IP disturbances, during which the solar wind speed decreased down to or below $\simeq$ 400 km s$^{-1}$.
% and the solar wind was observed to decrease again to or below 400 km s$^{-1}$ after the HSS transit. As an example, 
One isolated HSS was observed, for instance, during the BR 2494 between June 4 and June 10, 2016 (see Fig. 2). 
  In  Fig. 7 solar wind speed variations  above 400 km s$^{-1}$ from the beginning through the dip  of each  GCR flux depressions are shown with the corresponding flux decrease amplitudes.  In Fig. 8 twelve solar wind speed decreases down to 400 km s$^{-1}$ and associated GCR flux recovery amplitudes are shown. In Fig. 7 (8) the solar wind speed and the GCR flux variations must be intended positive (negative) and negative (positive), respectively.   
%The average characteristics of recurrent  GCR flux depressions  in space are reported in table 1. 
%The intensity of the same variations observed in space and on Earth with polar neutron monitors are compared in Figures 1 and 2. 
%The average value of GCR flux short-term variations  gathered in space in 2016-2017 5.1$\pm$2.5\% spanning from 1.6\% through 12.3\%. The average value of the depression intensities observed with neutron monitors is 1.8$\pm$1.6\%. 
%In (ref atmosphere) we have shown that the maximum GCR flux variations ranged from  -7\% through +8\% with respect to average value during the BR 2496 above 70 MeV n$^{-1}$, due to two Forbush decreases associated with the passage of Coronal Mass Ejections on July 20 and August 2. All short-term depressions are energy-dependent.  
%The maximum percent change of the GCR flux observed with LPF in space during recurrent GCR flux variations was of 12\% as it possible to infer from Figs. 3 and 4.
%In Armano et al., 2018 it was found that below 400 km s$^{-1}$ and with magnetic field below 10 nT the flux of GCR is not depressed with respect  to the maximum value observed during each BR. 
%However,  even when the mentioned interplanetary parameters present similar values the cosmic-ray variations may appear completely different. The behaviour of the GCR flux follows from the effects of 
%Unfortunately, 
%the effects on the GCR flux when the interplanetary parameter values are higher than these values appear different on the basis of subsequent interplanetary structures. 
On average, at the passage of individual HSS, the GCR flux is found to decrease  by 4.4\%  with average  solar wind speed increases of 226 km s$^{-1}$ above 400 km  s$^{-1}$. The most intense GCR intensity  decreases are  detected at the passage of IP disturbances if the GCR flux lies  in the range 5\%-7\% above the average value observed during the corresponding BR.  The smallest decreases are observed  at the transit of IP structures in case the GCR flux is already depressed.  When the solar wind speed stops increasing  but remains above
500 km s$^{-1}$ the cosmic-ray flux does not recover. 
%In this last case, the GCR flux is found to recover $>$ 3\% of their intensity. However, 
The recovery phases are characterized by average GCR intensity increases of 5.3\%  for mean solar wind speed decreases of 282 km s$^{-1}$. %above 400 km s$^{-1}$. 

In Fig. 9, the GCR depression commencement observed  on November 21 during the BR 2500 is modulated by an isolated HSS.\\ In order to estimate the pre-decrease LPF proton-dominated differential flux and the differential flux at the dip of the depression  on November 26-29, the proton AMS-02 measurements averaged during the same BR are considered first \citep{aguilar2018}. Solid stars in Fig. 10 represent the AMS-02 data and the  dot-dashed line was obtained from the  G\&A model for a solar modulation parameter of 385 MV/c (November 2016).  The  proton insterstellar spectrum by \citet{burger2000} was used for this calculation. The model outcomes are interpolated according to the following equation down to 70 MeV n$^{-1}$:

\begin{equation}
F(E)= A\ (E+b)^{-\alpha}\ E^{\beta}  \ \ \ {\rm particles\ (m^2\ sr\ s\ GeV\ n^{-1})^{-1}},
\label{equation2}
\end{equation}
 
\noindent found to well reproduce the model trend between tens of MeV and hundreds of GeV. 
\noindent In the above equation   $E$ is the particle kinetic energy per nucleon.  The
parameters $A$, $b$, $\alpha$ and $\beta$ are reported in Table 3 (see \cite{apj1} for details) for the dot-dashed line and for the top and bottom continuous lines of  Fig. 10.  The parameter $A$  represents the spectrum normalization while the parameter  $b$ determines the spectrum modulation below a few GeV. The parameters $\alpha$ and $\beta$ set the spectrum slope at high energies. By increasing the parameter $b$ the spectrum maintains the same spectral index at high energies  but is modulated below a few GeV. It is worthwhile to point out that this parameterization is used only for the purpose of reproducing the trend of the experimental data and for the G\&A model outcome parameterization. The parameter values are set on this basis only and no physical meaning must the assigned. The same approach was used to study the energy dependendence of FDs in \citet{apj1,apj2}. The continuous lines represent the GCR flux on November 20, 2016 (top line) and on November 26-29, 2016 (bottom line). These differential proton fluxes were obtained by taking into account that on November 20 the LPF GCR proton dominated integral flux was observed to be  +5.5\% above the  value averaged during the BR 2500   and on November 26-29,  3\% below. The proton flux at the dip of the depression appears modulated mainly below 2 GeV,
as it results from  Fig. 11 where it can be observed that polar NMs above 11 GeV  show a variation of about 1\%. 
Differential proton-dominated fluxes are inferred, and consequently the parameter $b$ is set, by requiring that the integral fluxes above 70 MeV and 11 GeV are consistent with the LPF and NMs measurements.
%at the peak and in the plateau region of the depression started on vvvvv, respectively.
%It is pointed out that despite integral fluxes differ by a few percent, 
%even if the top and bottom curves integral fluxes are +5.5\% and -3\% with respect to the average value (dot-dashed line), 
%the proton pre-decrease differential flux at 70 MeV differ by 16\% with respect to that at the dip. 
%In Fig. 11 it can be observed that polar NMs  show a variation of about 1\%.
These observations mark a difference with GCR flux modulation during FDs  of similar intensities observed with LPF \citep{apj2} for which polar and high latitude neutron monitors 
showed a variation of 4\% and  2\%, respectively. It is pointed out that the uncertainty of neutron monitor measurements on hourly averaged data is well below 1\%. In Figs. 12 and 13 the  energy dependence of GCR flux variations show that the  flux does not recover until the very last few days of the BR 2503  while neutron monitor observations do not show any evident variation. 
 
The GCR flux remains close to its average value estimated during  an entire BR either if  the solar wind speed remains below  400 km s$^{-1}$  or well   above 500 km s$^{-1}$. In the first case the GCR flux is at its maximum value during the studied BR. In the second, the GCR flux is depressed. We can assess that when  a large sample of GCR short-term variation evolutions down to a few tens of MeV  will be observed in space  on board the future space interferometers for gravitational waves and, possibly, other instruments before the LISA launch, more will be known about differences in the energy dependence of recurrent and non recurrent GCR flux variations. 
These findings must be correctly taken into account to carry out precise estimates of instrument performance in space.
For instance, the role of GCR energy differential flux variations in generating spurious noise force due to the charging process of free-falling metal test-masses of the LISA interferometer represents the main concern about mission performance below 10$^{-3}$ Hz.  
 The ESA Next Generation Radiation Monitors \citep{ngrm} will be flown on each LISA spacecraft. This detector consists  of two units, one for 2-200 MeV  proton and and one for 100 keV-7 MeV  electron monitoring. This detector is   optimized for solar energetic particle event detection and short-term forecast. However, the detector performance in the energy range of cosmic rays is quite poor. It was recently proposed to fly on LISA a third particle detector unit, similar to that located on board of LPF. This third unit may consist of only one silicon detector of 2.8$\times$2.8 cm$^2$ area placed inside a copper box stopping particles below 70 MeV s$^{-1}$. This unit, possibly oriented with its viewing axis perpendicularly to the nominal Parker spiral direction, would allow to monitor hourly GCR short-term variations within 1\% uncertainty.

%associated with solar wind speeds $>>$ 400 km s$^{-1}$ and IP magnetic field intensities larger than 10 nT  observed with LPF and recurrent geomagnetic activity are discussed. 
%GCR  flux recurrent short-term variations and  recurrent geomagnetic storms are typically of weak to moderate in intensity  
%Therefore, GCR recurrent depressions observed in  space are optimum proxies to forecast recurrent geomagnetic activity for space weather applications.
% with respect to using neutron monitor data as it was attempted in ref, due to the energy dependence of short-term GCR flux variations and despite the recurrent geomagnetic storms are typically of weak to moderate in intensity.  

\section{Recurrent galactic cosmic-ray flux variations and geomagnetic activity in 2016-2017}  
In \cite{grim19} it was explored the possibility to use three weak FDs (July 20, 2016, August 2, 2016, May 27, 2017) observed on board LPF  as proxies for non-recurrent geomagnetic activity associated with the arrival at the Earth of the associated ICMEs. It was observed that FDs could have been used for forecasting  intense geomagnetic storms only when  a negative B$_z$ component (in Geocentric Solar Ecliptic (GSE) coordinate system) of the IP magnetic field reconnecting with the Earth's magnetic field  was assuming values smaller than -20 nT and the interplanetary magnetic field intensity was $>$ 20 nT. However, in \cite{telloni1}  it was shown that if the solar wind energy is large enough, geomagnetic activity is observed also when the B$_z$ of the IP magnetic field is  northern. To this purpose, recurrent depressions of the GCR flux observed with LPF and recurrent geomagnetic activity are discussed.

The association between corotating solar wind streams and weak to moderate recurrent geomagnetic activity was discussed in
\cite{tsurutani}. It is recalled here that a weak geomagnetic activity is correlated with the disturbance storm time (Dst) geomagnetic index ranging between -30 nT and -50 nT, while the geomagnetic activity is considered moderate in case the Dst varies between -50 nT and -100 nT. 
%The association among FDs and  intense geomagnetic activity in 2016-2017 was dicussed in  \cite{apj2,grim19} and found that FDs with amplitudes below 10\% are not good proxies for intense geomagnetic storms. 
Weak-moderate geomagnetic activity in 2016-2017 during the time LPF remained into orbit was observed when the B$_z$ component of the IP magnetic field presented values $<$ -12 nT. However, weak-moderate geomagnetic activity was also observed at the leading edge of  fast solar wind streams characterized by minimum speeds ranging between 550 km s$^{-1}$ and 650 km s$^{-1}$ and a region of enhanced plasma pressure  of at least 12 nPa regardless of the B$_z$ value. 
%In regions where the B$_z$ component of the IP magnetic field presents values $<$ -12 nT weak-moderate geomagnetic activity is also observed regardless of the solar wind speed and plasma pressure. 
Recurrent geomagnetic activity  was associated with passage of HSS from the BR 2491 through the BR 2494 on March 16-17, 2016 (Dst $\simeq$ -50 nT), April 13-14, 2016 (Dst $\simeq$ -55 nT), May 8-9, 2016  (Dst $\simeq$ -80 nT) and June 5-6, 2016 when the geomagnetic activity became weak (Dst $\simeq$ -40 nT). Analogously, recurrent geomagnetic activity was also observed on September 1-2, 2016 
(Dst $\simeq$ -60 nT), September 29, 2016 (Dst $\simeq$ -60 nT) and  October 25, 2016 (Dst $\simeq$ -50 nT) from BR 2497 through BR 2499 . The weak to moderate geomagnetic activity is always associated with GCR flux recurrent variations in L1. It must be pointed out, however, that  GCR flux depressions due to the interplay of closely spaced HSS ( $<$ 1 day) characterized by   solar wind speed $>$ 400 km s$^{-1}$ cannot trigger geomagnetic activity because of a small solar wind energy with typical flow pressure of just a few nPa \citep{telloni1}. 

 GCR short-term variation observations in L1 not only  help in disentangling long and short term variations of the GCR flux but also represent a precious proxy of weak to moderate geomagnetic activity.

\begin{figure}
 \plotone{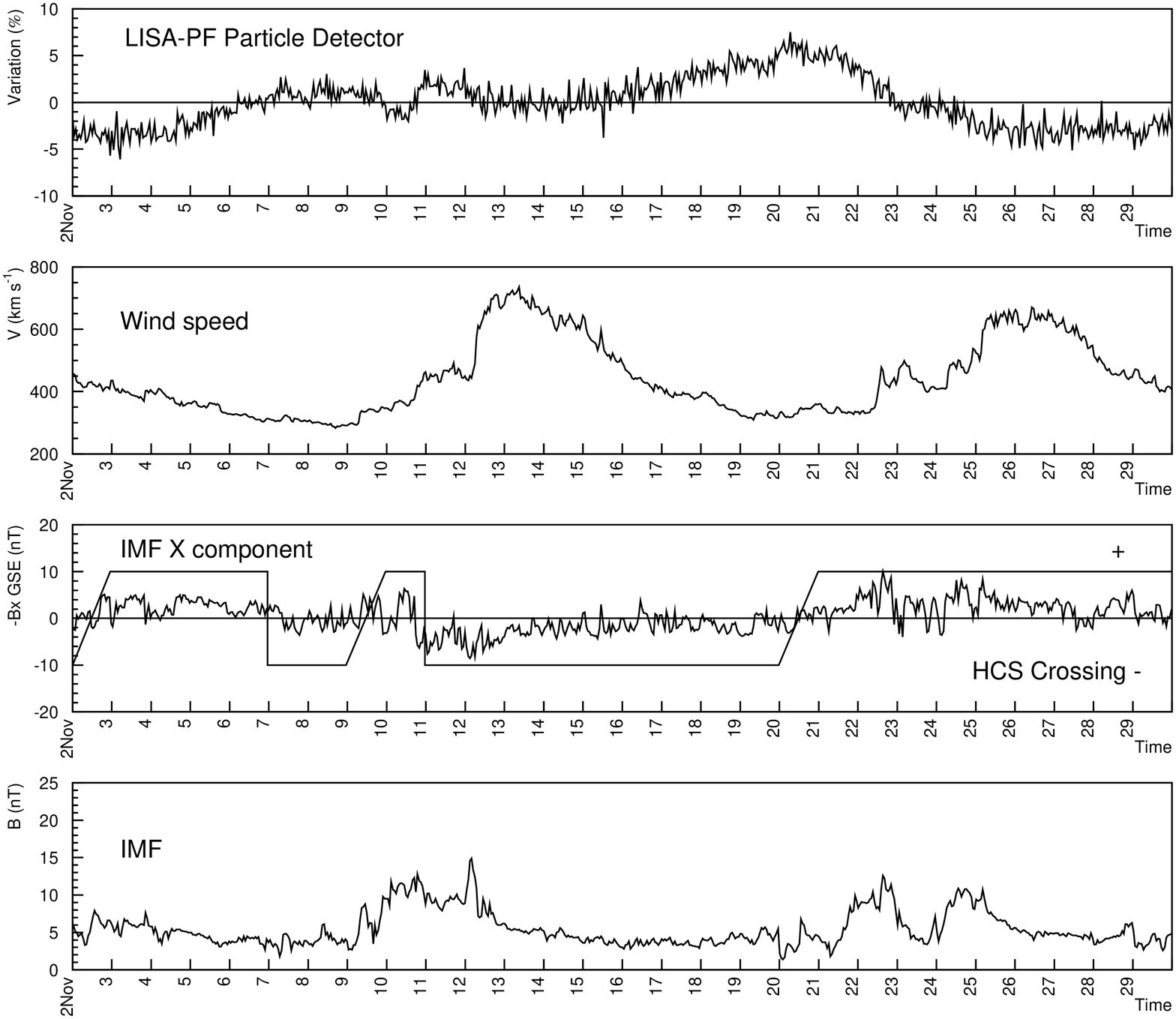}
  \caption{Same as Fig. 1 for the BR 2500 (November 2, 2016 - November 28, 2016).}
  \label{figure1}
\end{figure}

\begin{figure}
  \plotone{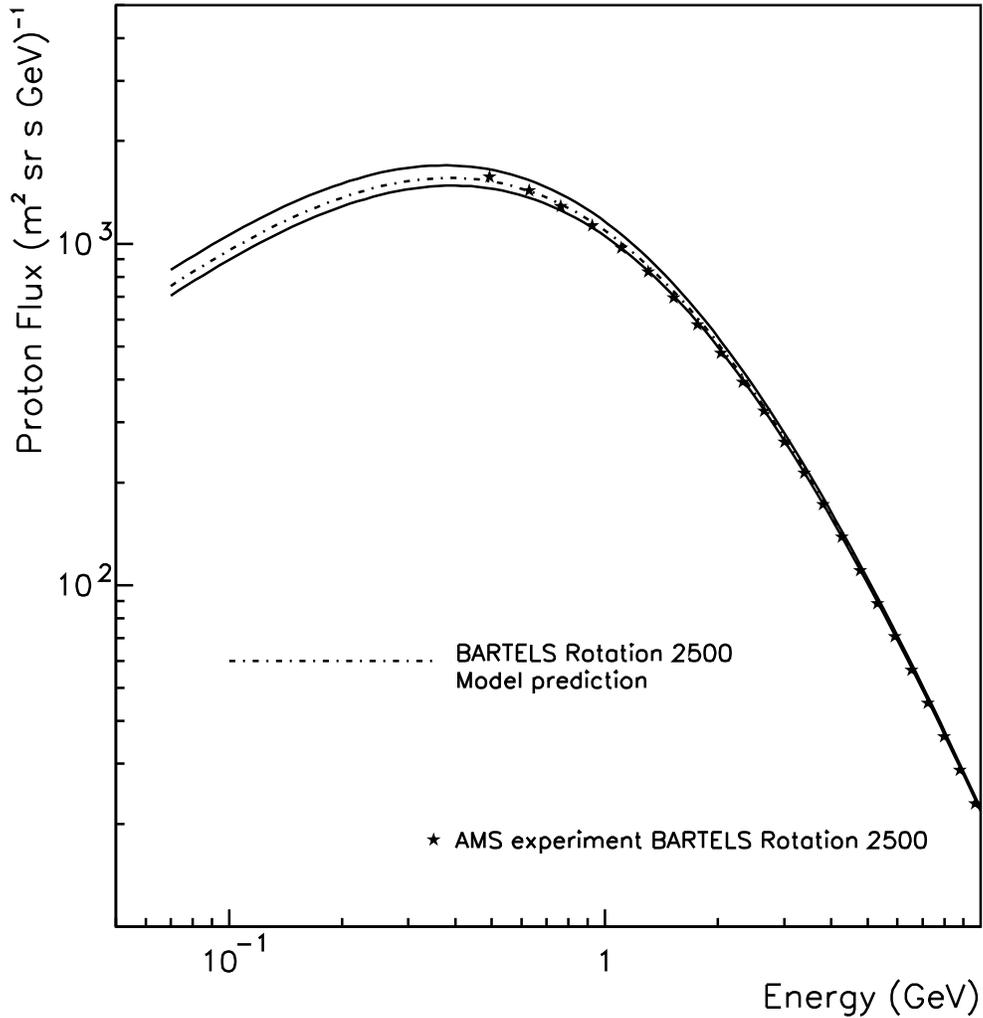}
  \caption{Energy dependence of the GCR proton dominated flux depression commenced on  November 20, 2016 (top continuous line)  during the BR 2500 on board LPF. The bottom continuous line represents the same flux at the maximum of the depression on  November 26-29 2016. The dot-dashed line shows the average GCR proton flux during the BR 2500 resulting from the interpolation (equation 1) of the AMS-02 experiment data (solid stars).}
  \label{figure1}
\end{figure}

\begin{figure}
  \plotone{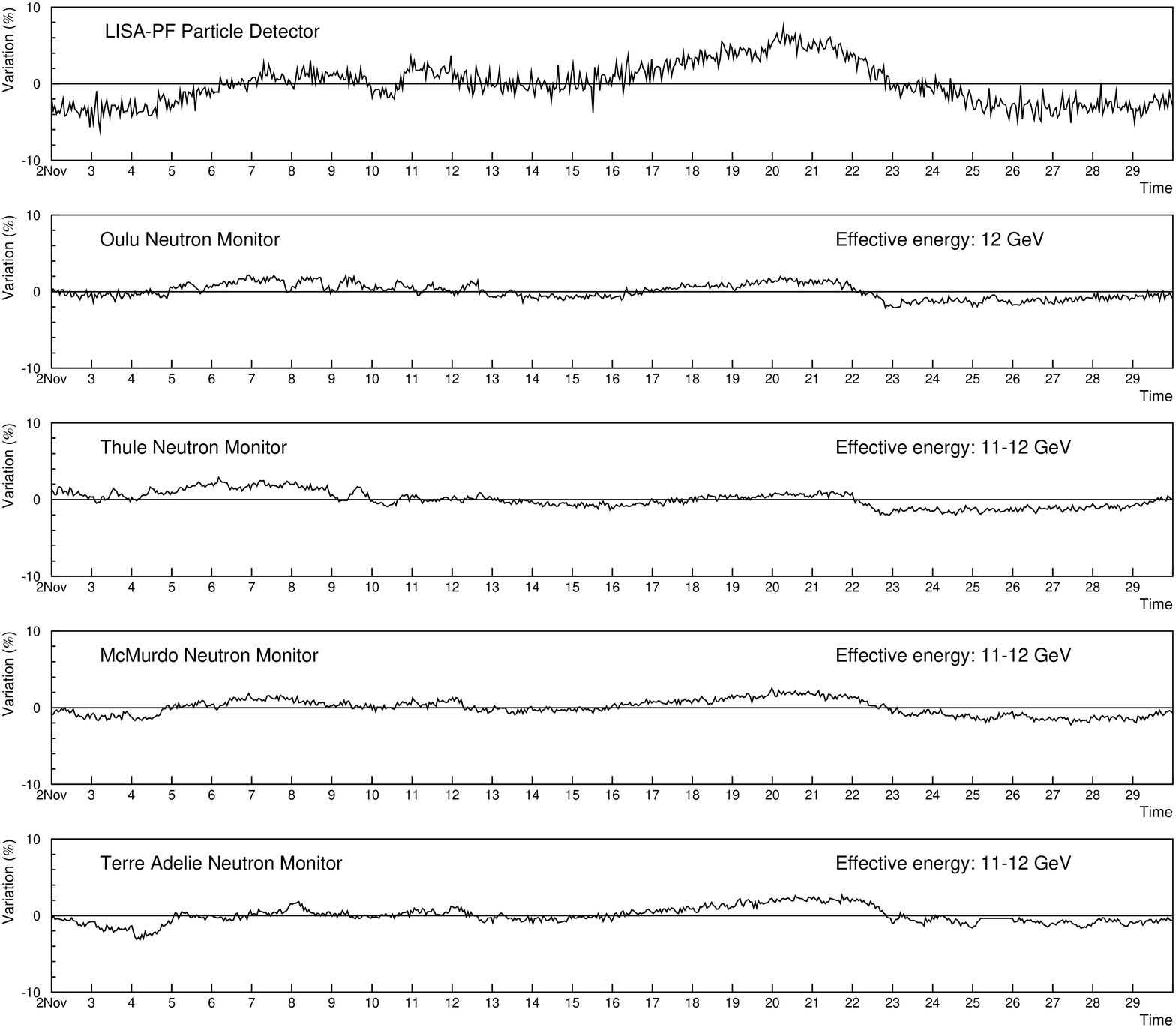}
  \caption{Same as Fig. 2 for the BR 2500 (November 2, 2016 - November 29, 2016).}
  \label{figure1}
\end{figure}

\begin{figure}
 \plotone{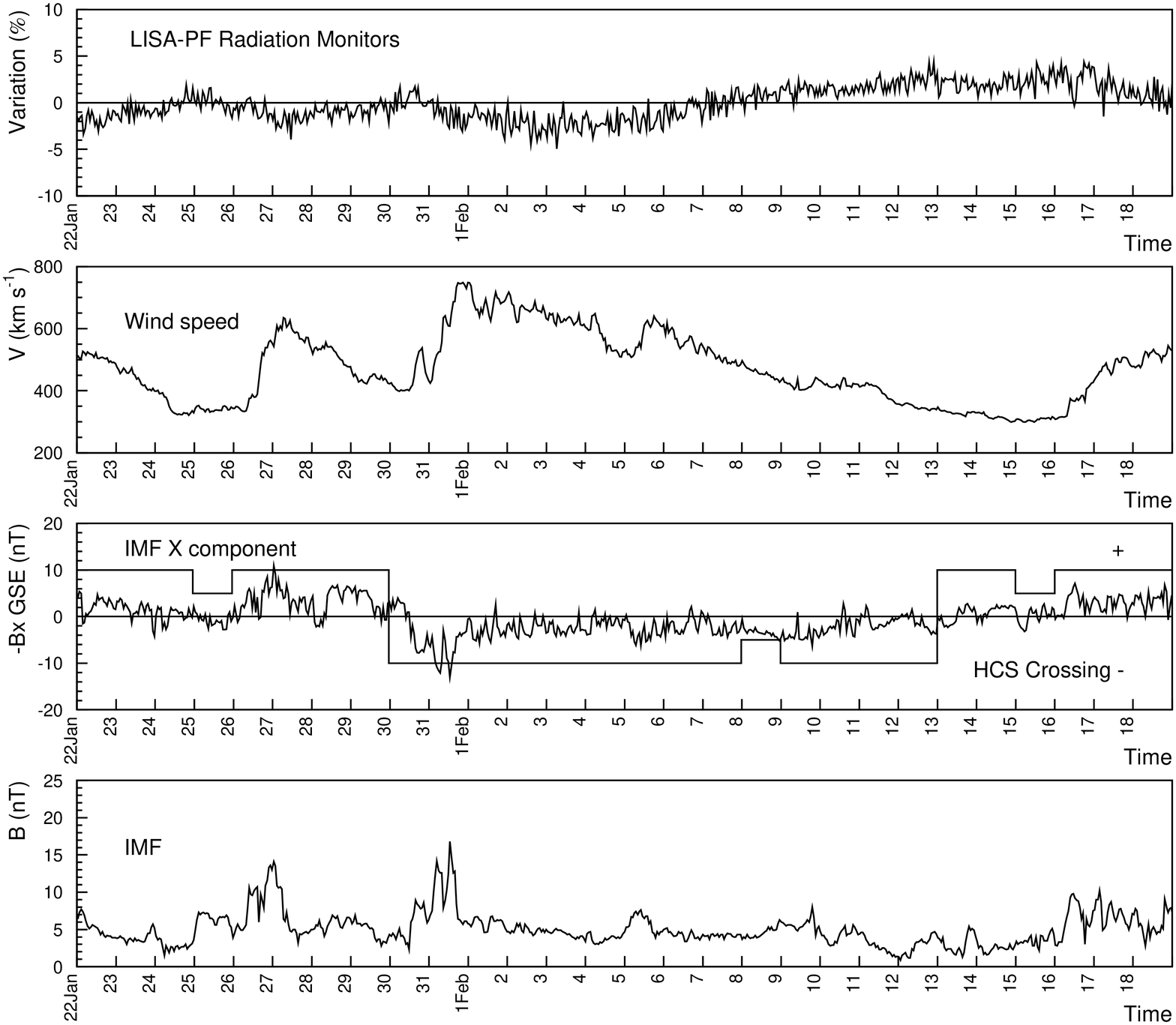}
  \caption{Same as Fig. 1 for BR 2503 (January 22, 2016- February 17, 2016).}
  \label{figure1}
\end{figure}

\begin{figure}
    \plotone{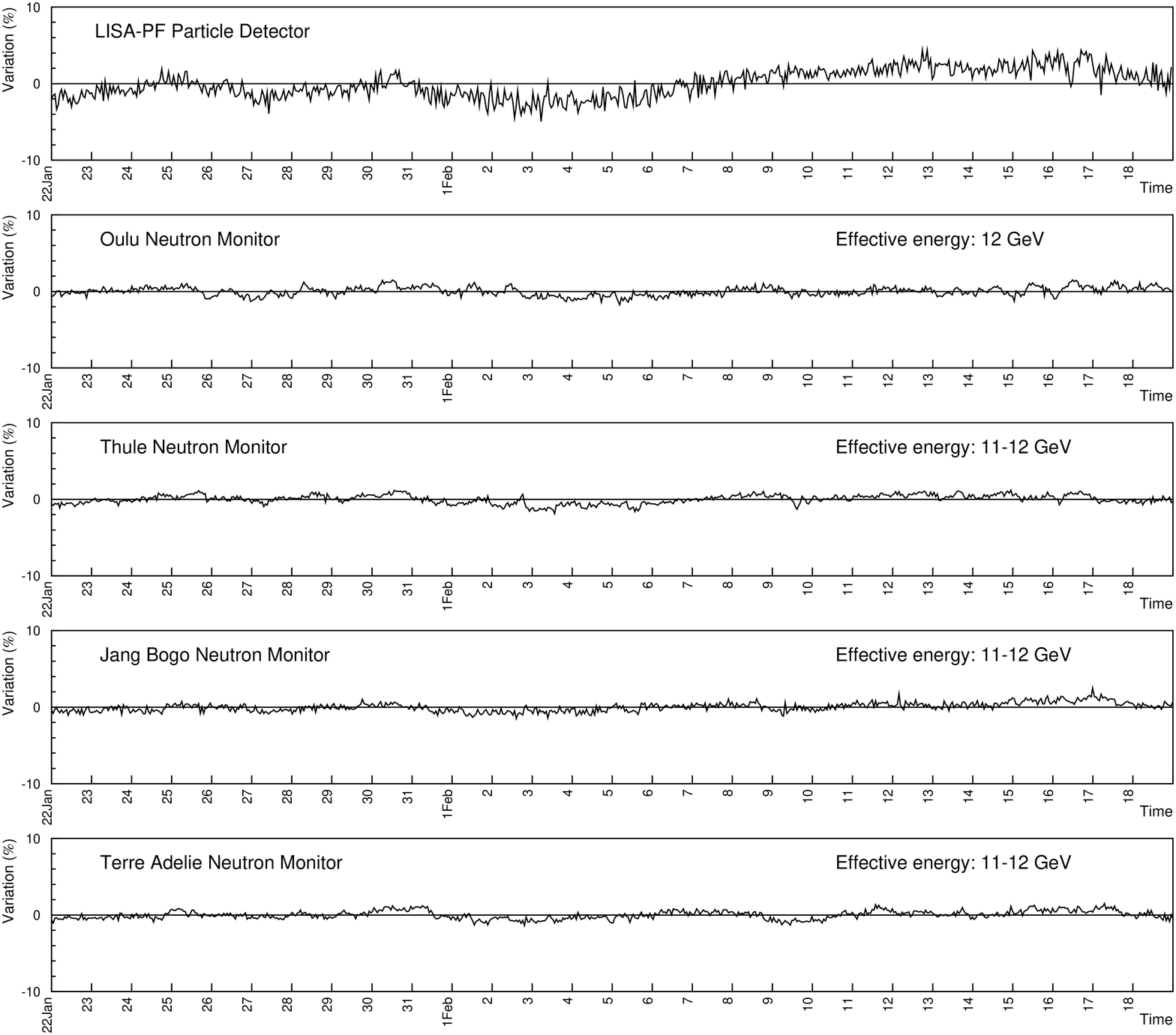}
  \caption{Same as Fig. 2 for BR 2503 (January 22, 2016- February 17, 2016). }
  \label{figure1}
\end{figure}

\begin{table}
\centering
\caption{\label{table1} Parameterizations of proton energy spectra averaged over the BR 2500,  before and at the dip of the depression started on November 22, 2016.}
\begin{tabular}{lllll}%{@{}*{9}{l}}                                                                                                                                   
\hline
\hline
&  $A$  &  $b$ &   $\alpha$  &  $\beta$  \\
\hline
AMS-02 proton flux  & 18000 & 1.15 & 3.66 & 0.92\\
Pre-decrease proton flux  & 18000 & 1.115 & 3.66 & 0.92\\
Proton flux at the depression dip  & 18000 & 1.172 & 3.66 & 0.92\\
\hline
\end{tabular}
\end{table}

\section{Conclusions}

The energy differential flux  of GCRs during the evolution of individual recurrent short-term variations has been studied by comparing data
 gathered in deep space with the ESA mission LPF, with  cosmic-ray dedicated experiments in space and with polar NMs. 
The passage of HSS of similar characteristics are associated with  GCR depression amplitudes differing by more than a factor of two depending if the 
particle flux was close to minimum or maximum values during each BR  before the passage of modulating HSS.
An almost constant cosmic-ray flux is observed when the solar wind speed remains below 400 km s$^{-1}$ or  
above 500 km s$^{-1}$. In the first case the GCR flux is at its maximum value for the period of the solar cycle and solar polarity. In the second case, the flux is steadily depressed by the passage of closely spaced  HSS.
The GCR flux during recurrent short-term depressions presented  maximum amplitudes of about 12\%  in 2016-2017 and resulted   
modulated mainly below 2 GeV while during FDs of the same amplitude, and energy dependence of several percent  was observed above 10 GeV. 
   
GCR flux depressions in L1 are  good proxies to monitor weak to moderate geomagnetic activity being both processes associated with the passage
 of corotating HSS. The only exception is associated with the transit of several HSS when the solar wind speed does not decrease at 400 km s$^{-1}$ and the GCR flux is already depressed. In this case, the formation of regions of enhanced pression $>$ 12 nPa is impeded and consequently the solar wind energy is not large enough to disturb the geomagnetic field regardless of the B$_z$ value.

\vspace*{0.5cm}

\subsection*{\bf Acknowledgements}
\footnotesize{}

%%\begin{table}
%%\caption{\label{tabone}Parameterization of the primary cosmic-ray fluxes
%% at solar minimum ($A_m$,$B_m$,$\alpha_m$,$\beta_m$) and solar maximum ($A_M$,$B_M$,$\alpha_M$,$\beta_M$) }
%%%%\begin{indented}
%%%%\lineup
%%%%\item[]\begin{tabular}{@{}*{7}{l}}
%%%%\br                              
%%%%$\0\0A$&$B$&$C$&\m$D$&\m$E$&$F$&$\0G$\cr 
%%%\mr
%%%\0\023.5&60  &0.53&$-20.2$&$-0.22$ &\01.7&\014.5\cr
%%%\0\039.7&\-60&0.74&$-51.9$&$-0.208$&47.2 &146\cr 
%%%\0123.7 &\00 &0.75&$-57.2$&\m---   &---  &---\cr 
%%%3241.56 &60  &0.60&$-48.1$&$-0.29$ &41   &\015\cr 
%%%\br
%%%%\end{tabular}
%%%%\end{indented}
%%%\end{table}

The LISA Pathfinder cosmic-ray data can be downloaded from \url{https://www.cosmos.esa.int/web/lisa-pathfinder-archive/home}. 
Sunspot number data were gathered from \url{http://www.sidc.be/silso/home}. 
Data from ACE and Wind experiments are taken from the NASA-CDAWeb website. 
The ICME catalog appears in \url{http://www.srl.caltech.edu/ACE/ASC/DATA/level3/icmetable2.htm}. 
NMs data are gathered from \url{www.nmdb.eu}. The authors thank the PIs of the NM network.

%\bibliography{biba_1_m}

\bibliography{biba20sett20}{}
\bibliographystyle{aasjournal}



\end{document}